\documentclass[aip,twocolumn,superscriptaddress]{revtex4}
   \usepackage{graphicx}
   \usepackage{amssymb}
   \usepackage{amsmath}
   \usepackage{subfigure}
   \usepackage{graphpap}
   \usepackage{dcolumn} 
   \usepackage{bm}
   \usepackage{color,calc}
   \usepackage{longtable}
   \usepackage{hyperref}
   \newcommand{\bra}{\langle}
   \newcommand{\ket}{\rangle}
   
   \renewcommand{\b}[1]{\ensuremath{\mathbf{#1}}}

\begin{document}
   %
\title{\textit{Ab initio} lifetime correction to scattering states for time-dependent electronic-structure calculations with incomplete basis sets}
   \author{Emanuele Coccia} \email{emanuele.coccia@nano.cnr.it}
   \affiliation{Dipartimento di Scienze Fisiche e Chimiche, 
                Universit\'a degli Studi dell'Aquila, via Vetoio, 67100 L'Aquila, Italy}
   \affiliation{Laboratoire de Chimie Th\'eorique, Universit\'e Pierre et Marie Curie, Sorbonne Universit\'es, CNRS, Paris, France}
   \author{Roland Assaraf} \email{assaraf@lct.jussieu.fr}
   \affiliation{Laboratoire de Chimie Th\'eorique, Universit\'e Pierre et Marie Curie, Sorbonne Universit\'es, CNRS, Paris, France}
   \author{Eleonora Luppi}\email{eleonora.luppi@upmc.fr}
   \affiliation{Laboratoire de Chimie Th\'eorique, Universit\'e Pierre et Marie Curie, Sorbonne Universit\'es, CNRS, Paris, France}
   \author{Julien Toulouse}   \email{julien.toulouse@upmc.fr}
   \affiliation{Laboratoire de Chimie Th\'eorique, Universit\'e Pierre et Marie Curie, Sorbonne Universit\'es, CNRS, Paris, France}

\date{June 19, 2017}
\begin{abstract}
We propose a method for obtaining effective lifetimes of scattering electronic states for avoiding the artificially confinement of the wave function due to the use of incomplete basis sets in time-dependent electronic-structure calculations of atoms and molecules. In this method, using a fitting procedure, the lifetimes are extracted from the spatial asymptotic decay of the approximate scattering wave functions obtained with a given basis set. The method is based on a rigorous analysis of the complex-energy solutions of the Schr\"odinger equation. It gives lifetimes adapted to any given basis set without using any empirical parameters. The method can be considered as an \textit{ab initio} version of the heuristic lifetime model of Klinkusch {\it et al.} [J. Chem. Phys. \textbf{131}, 114304 (2009)]. The method is validated on the H and He atoms using Gaussian-type basis sets for calculation of high-harmonic-generation spectra.
\end{abstract}

\maketitle

\section{Introduction}

Motivated by experimental advances in attosecond science~\cite{cor07,kra09,gal12,atto,paz15,cal16,RamLeoNeu-ARPC-16}, there is currently a lot of interest in developing time-dependent electronic-structure computational methods for studying laser-driven electron dynamics in atomic and molecular systems (see, e.g., Ref.~\onlinecite{HocHinBon-EPJST-14}). Examples of such methods include time-dependent density-functional theory (TDDFT)~\cite{RunGro-PRL-84}, time-dependent Hartree-Fock (TDHF)~\cite{Kul-PRA-87b}, multiconfiguration time-dependent Hartree-Fock (MCTDHF)~\cite{CaiZanKitKocKreScr-PRA-05}, time-dependent configuration interaction (TDCI)~\cite{KraKlaSaa-JCP-05}, and time-dependent coupled cluster~\cite{Kva-JCP-12}. These methods involve orbitals which are often expanded on basis functions such as Gaussian-type functions~\cite{KraKlaSaa-JCP-05,Krause:2007dm,lupp+12mol,lupp+13jcp,coccia16a,white15,coccia16b}, and an important question is whether the continuum scattering states which are explored at high laser intensity, e.g. in the high-harmonic generation (HHG) process, are sufficiently well described.

The description of the continuum scattering states can be much improved by using specially designed Gaussian-type basis sets, such as the one proposed by Kaufmann {\it et al.}~\cite{kauf+89physb}, as demonstrated recently in Refs.~\onlinecite{coccia16a,coccia16b}. However, even with these basis sets, only an incomplete discrete set of scattering states which decay too fast away from the nucleus is obtained, with the consequences that ionization processes cannot be properly described and the time-dependent wave function undergoes artificial reflections. These problems can be alleviated with \textit{ad hoc} lifetime models as proposed by Klinkusch {\it et al.}~\cite{Klinkusch:2009iw} or Lopata {\it et al.}~\cite{lop13} which introduce an imaginary part to the energy of each scattering state. This has the effect of partially absorbing the time-dependent wave function which limits artificial reflections and simulates ionization. However, these lifetime models have the disadvantage of being empirical and of depending on adjustable parameters. Alternative methods to these lifetime models include approaches using a complex-absorbing potential (CAP)~\cite{GolSho-JPB-78,LefWya-JCP-83,KosKos-JCC-86,MugPalNavEgu-PR-04,GreHoPabKamMazSan-PRA-10}, an absorbing mask function~\cite{kra92}, or exterior-complex scaling~\cite{MccStrWis-PRA-91,he07,tao09,scr10,TelSosRozChu-PRA-13} to absorb the time-dependent wave-function beyond a certain distance, but all these techniques also inevitably imply some empiricism in the choice of the involved parameters. In the present work, we develop an \textit{ab initio} lifetime correction to scattering states based on a rigorous analysis of the complex-energy solutions of the Schr\"odinger equation. This \textit{ab initio} correction gives lifetimes adapted to each particular incomplete basis set without any free parameters. 

More specifically, we start from the exact complex-energy solutions of the Schr\"odinger equation of a hydrogen-like atom, obtained by relaxing the boundary conditions on the wave function, making the Hamiltonian a non-Hermitian operator. For a complex-energy state, we show that the value of the corresponding lifetime is encoded in the spatial asymptotic behavior of the associated wave function. We thus propose, for a given Gaussian-basis set, to extract an effective lifetime associated with an approximate scattering state of real energy by matching its spatial asymptotic decay with the one of the exact complex-energy wave function having the same real part of the energy. In practice, this is done with a fit of the spatial asymptotic decay of each scattering wave function and leads to parameter-free lifetimes for the one-electron scattering states, which compensate for the incompleteness of the basis set in time-dependent calculations. We then show how the procedure can be extended to many-electron atoms and molecules to define lifetimes for $N$-electron scattering states used in TDCI calculations. Interestingly, the lifetimes defined in the heuristic model of Klinkusch {\it et al.}~\cite{Klinkusch:2009iw} are recovered as simple approximations of our lifetimes, which clarifies the theoretical grounds of this model.

The paper is organized as follows. In Section~\ref{theory}, we present in detail the theory of our \textit{ab initio} lifetime correction. In Section~\ref{details}, we give computational details for the tests performed on the H and He atoms. In Section~\ref{results}, we give and discuss the results. In particular, we show the effect of using the \textit{ab initio} lifetime correction for calculating HHG spectra and we compare with the heuristic lifetime model. Finally, Section~\ref{conclusion} contains our conclusions. Unless otherwise stated, Hartree atomic units are used throughout the paper.

\section{Theory}
\label{theory}

\subsection{Schr\"odinger equation for a hydrogen-like atom with complex energies}
\label{Schrodinger}

Consider the time-independent Schr\"odinger equation for a hydrogen-like atom (with a nuclear charge $Z$)
\begin{equation}
\left( -\frac{1}{2} \nabla_\b{r}^2 - \frac{Z}{r} \right) \psi(\b{r}) = E \psi(\b{r}),
\end{equation}
with a possibly complex energy $E$ and the associated electronic wave function $\psi(\b{r})=R(r) Y_{\ell}^{m}(\theta,\phi)$ written as the product of a radial part $R(r)$ and a spherical harmonics $Y_{\ell}^{m}(\theta,\phi)$. The radial part is determined by the equation
\begin{equation}
R''(r) + \frac{2}{r} R'(r) + \left( - \frac{\ell(\ell+1)}{r^2} + \frac{2Z}{r} +2 E \right) R(r) = 0,
\label{radialeq}
\end{equation}
for a given angular momentum $\ell$. The general solution of Eq.~(\ref{radialeq}), without imposing any boundary conditions, is (as found, e.g., with {\sc Mathematica}~\cite{Math10-PROG-15}; see also Ref.~\onlinecite{Mah-BOOK-09})
\begin{eqnarray}
R(r) &=& c_1 R_1(r) + c_2 R_2(r),
\label{solR}
\end{eqnarray}
where $c_1$ and $c_2$ are two arbitrary complex constants, and 
\begin{eqnarray}
R_1(r) &=& L\left(\nu,2\ell+1,2\sqrt{-2E} \; r\right) 
r^\ell e^{-\sqrt{-2E} \; r},
\label{solR1}
\end{eqnarray}
and
\begin{eqnarray}
R_2(r) &=& U\left(-\nu, 2\ell+2,2\sqrt{-2E} \; r\right) 
r^\ell e^{-\sqrt{-2E} \; r},
\label{solR2}
\end{eqnarray}
where $\nu=Z/\sqrt{-2E} -\ell-1$.~\cite{lifetimes-note2} In these expressions, $L$ is the generalized Laguerre function and $U$ is the Tricomi confluent hypergeometric function, both defined for possibly complex arguments. The function $R_1(r)$ is always finite at $r=0$
\begin{eqnarray}
|R_1(r=0)| < \infty,
\end{eqnarray}
and, for generic values of the complex energy $E$, its asymptotic behavior for $r\to\infty$ is
\begin{eqnarray}
R_1(r) &\underset{r\to\infty}{\sim}& \frac{1}{\Gamma(-\nu)}\frac{e^{\sqrt{-2E} \; r}}{r^{1+Z/\sqrt{-2E}}},
\label{R1inf}
\end{eqnarray}
where $\Gamma(z)$ is the gamma function. For generic values of $E$, the function $R_2(r)$ diverges at $r=0$ as
\begin{eqnarray}
R_2(r) &\underset{r\to0}{\sim}& \frac{1}{\Gamma(-\nu)} \frac{1}{r^{\ell+1}},
\label{R2r0}
\end{eqnarray}
while its asymptotic behavior for $r\to\infty$ is
\begin{eqnarray}
R_2(r) &\underset{r\to\infty}{\sim}& \frac{e^{-\sqrt{-2E} \; r}}{r^{1-Z/\sqrt{-2E}}}.
\label{R2inf}
\end{eqnarray}

Let us consider first the case of real and negative energies, $E = \varepsilon <0$. In this case, the function $R_1(r)$ generally diverges as $e^{\sqrt{-2 \varepsilon} \; r}$ for $r\to\infty$ [Eq.~(\ref{R1inf})] and the function $R_2(r)$ generally diverges as $1/r^{\ell+1}$ at $r=0$ [Eq.~(\ref{R2r0})]. However, when $\nu$ is a negative or zero integer, i.e. for the discrete energy values $\varepsilon = -Z^2/(2n^2)$ where $n$ is a positive integer with $n \geq \ell+1$, the prefactor $1/\Gamma(-\nu)$ goes to $0$ in Eq.~(\ref{R1inf}) and the divergence of the function $R_1(r)$ is avoided. In fact, Eq.~(\ref{R2r0}) has the same prefactor $1/\Gamma(-\nu)$ and the divergence of the function $R_2(r)$ is also avoided for these discrete energy values, as less well known~\cite{OthMonMar-ARX-16}. For these particular energy values, it turns out that the functions $R_1(r)$ and $R_2(r)$ both become proportional to the familiar associated Laguerre polynomials, so that one is free to choose any linear combination of $R_1(r)$ and $R_2(r)$ to obtain proper (finite and normalizable) eigenfunctions. This case corresponds to the discrete bound states.

Consider now the case of real and positive energies, $E = \varepsilon > 0$. In this case, it can be seen from Eqs.~(\ref{R1inf}) and~(\ref{R2inf}) that $|R_1(r)|$ and $|R_2(r)|$ both behave asymptotically as $1/r$ [multiplied by an oscillatory cosine term for $R_1(r)$ due to an additional term not shown in Eq.~(\ref{R1inf})], but the function $R_2(r)$ diverges as $1/r^{\ell+1}$ at $r=0$ [Eq.~(\ref{R2r0})]. One thus has to set $c_2=0$ to obtain finite (but not normalizable) eigenfunctions. This case corresponds to the continuum scattering states.

Finally, let us consider complex energies, $E = \varepsilon - i \gamma/2$, with a positive real part, $\varepsilon > 0$, and a negative imaginary part, $-\gamma/2 < 0$. The corresponding states are interpreted as decaying states with a finite lifetime $\tau = 1/\gamma$ in the sense that the time-evolved wave function $\psi(\b{r},t) = e^{-i (\varepsilon-i\gamma/2) t} \psi(\b{r},0)$ has a survival probability which decays in time as $|\psi(\b{r},t)|^2/|\psi(\b{r},0)|^2=e^{-\gamma t}$. For determining the asymptotic behavior of the radial functions $R_1(r)$ and $R_2(r)$, it is convenient to define the free-electron momentum
\begin{equation}
k = \sqrt{2E}= k_\text{r} + i k_\text{i},
\end{equation}
with a positive real part
\begin{equation}
k_\text{r}=\sqrt{\frac{2\varepsilon+\sqrt{4\varepsilon^2+\gamma^2}}{2}},
\end{equation}
and a negative imaginary part
\begin{equation}
k_\text{i}=-\sqrt{\frac{-2\varepsilon+\sqrt{4\varepsilon^2+\gamma^2}}{2}}.
\label{ki}
\end{equation}
Using $\sqrt{-2E}= i k = i k_\text{r} + |k_\text{i}|$ in Eq.~(\ref{R1inf}), we thus find that the function $R_1(r)$ diverges exponentially for $r\to\infty$ as
\begin{eqnarray}
|R_1(r)| &\underset{r\to\infty}{\sim}& \frac{e^{|k_\text{i}| r}}{r^{1+Z |k_\text{i}|/|k|^2}},
\end{eqnarray}
so it cannot be considered as a proper eigenfunction but it is a kind of resonant state in which the electron ``escapes'' at infinity (see, e.g., Refs.~\onlinecite{HatSasNakPet-PTP-08,Moi-BOOK-11}). On the contrary, using Eq.~(\ref{R2inf}), we see that the function $R_2(r)$ goes exponentially to zero for $r\to\infty$,
\begin{eqnarray}
|R_2(r)| &\underset{r\to\infty}{\sim}& \frac{e^{-|k_\text{i}| r}}{r^{1-Z |k_\text{i}|/|k|^2}},
\label{R2infcomplexenergy}
\end{eqnarray}
but still diverges as $1/r^{\ell+1}$ at $r=0$ [Eq.~(\ref{R2r0})]~\cite{lifetimes-note1}. Therefore, nor can it be considered as a proper eigenfunction and again it may be thought of as a kind of resonant state in which the electron ``escapes'' at the position of the nucleus. Note that on the space of such diverging functions, the Hamiltonian is not a self-adjoint operator, which is why the eigenvalues can be complex.

From the above analysis, we thus see that one useful property of a hydrogen-like electronic state with a complex energy $E = \varepsilon - i \gamma/2$ and $c_1=0$ is that the inverse lifetime $\gamma$ of the state can be obtained from [after inverting Eq.~(\ref{ki})]
\begin{eqnarray}
\gamma= 2|k_\text{i}| \sqrt{2\varepsilon+|k_\text{i}|^2},
\label{Gamma}
\end{eqnarray}
where $|k_\text{i}|$ can be extracted from the asymptotic behavior of the radial function
\begin{eqnarray}
|k_\text{i}| = -\lim_{r\to\infty}\frac{\ln |R(r)|}{r}.
\end{eqnarray}
In particular, in the case of a scattering state for which $|R(r)| \sim 1/r$ as $r\to\infty$, we correctly obtain $|k_\text{i}|=0$ and $\gamma=0$. For physical interpretation of Eq.~(\ref{Gamma}), we note that $1/(2|k_\text{i}|)$ can be thought of as a measure of the spatial extension of the state and $\sqrt{2\varepsilon+|k_\text{i}|^2}=k_\text{r}$ can be interpreted as the velocity of the escaping electron~\cite{KlaGil-AQC-12}. In the following, we exploit this link between the spatial asymptotic decay of the state and its lifetime to formulate an \textit{ab initio} lifetime correction to scattering states for compensating the use of incomplete basis sets. 


\subsection{Ab initio lifetime correction to one-electron scattering states for incomplete basis sets}
\label{lifetimeoneelec}

We still first consider a one-electron hydrogen-like atom. In standard quantum chemistry programs, the Schr\"odinger equation is solved using an incomplete Gaussian-type basis set. For each state $p$, the radial function is expanded on $M$ basis functions $\{\chi_\mu (r)\}$
\begin{equation}
R_p(r) = \sum_{\mu=1}^M c_{\mu,p} \; \chi_\mu (r),
\end{equation}
where $c_{\mu,p}$ are the calculated orbital coefficients. Each basis function $\chi_\mu (r)$ of angular momentum $\ell_{\mu}$ is generally itself a contraction of $M_\mu$ primitive Gaussian-type basis functions
\begin{equation}
\chi_\mu (r) = \sum_{i=1}^{M_\mu} d_{i,\mu} \; r^{\ell_{\mu}} \; e^{-\alpha_{i,\mu} \; r^2},
\end{equation}
where $d_{i,\mu}$ and $\alpha_{i,\mu}$ are the (fixed) coefficients and exponents, respectively, of the $i^\text{th}$ primitive in the contraction. Obviously, in addition to the bound states with negative energies $\varepsilon_p <0$, discrete states with positive energies $\varepsilon_p >0$ are also obtained and they can be considered as approximations to the exact continuum scattering states. These approximate positive-energy states usually reproduce a number of oscillations of the exact scattering states, but they go to zero much faster than $1/r$ for large $r$ due to the limitation of the basis. When doing time-dependent calculations with this basis, this too fast decay of the approximate radial functions (and the fact that only a limited discrete set of states is obtained) artificially confines the electron around the nucleus, with the consequences that ionization processes cannot be properly described and the time-dependent wave function may undergo artificial reflections at the boundary of the space covered by the basis.

Clearly, beyond a large enough $r$, the approximate radial function $R_p(r)$ decays as $e^{-\alpha_\text{min} r^2}$ where $\alpha_\text{min}$ is the smallest exponent appearing in the expansion of $R_p(r)$. However, for an intermediate range of $r$, we have found that the envelope of $R_p(r)$ can be well described by the asymptotic behavior of the complex-energy state in Eq.~(\ref{R2infcomplexenergy}), i.e.
\begin{equation}
\text{envelope}[R_p(r)] \approx A_p \frac{e^{-B_p r}}{r^{C_p}} \;\;\;\text{for}\;\;\; r_\text{min} < r <r_\text{max},
\label{envRapprox}
\end{equation}
where $A_p$, $B_p$, $C_p$ are constants to be determined for each positive-energy state $p$. Therefore, we reinterpret the (envelope of the) approximate positive-energy state $R_p(r)$ as an approximation to a complex-energy state with spatial exponential decay, rather than an approximation to a real-energy scattering state with $1/r$ asymptotic behavior. Following Eq.~(\ref{Gamma}), we thus assign an inverse effective lifetime $\gamma_p$ to each such approximate state $R_p(r)$, obtained from the calculated energy of this state $\varepsilon_p >0$ and the decay exponent $B_p$
\begin{eqnarray}
\gamma_p = 2 B_p \sqrt{2\varepsilon_p+B_p^2}.
\label{Gamman}
\end{eqnarray}
Naturally, we extract $\gamma_p$ from the constant $B_p$ since it dominates the asymptotic behavior, but we note that $\gamma_p$ may also be extractable from $C_p$, according to Eq.~(\ref{R2infcomplexenergy}).

To obtain $B_p$, we fit the envelope of $R_p(r)$ for each state $p$. In principle, the envelope could be mathematically defined and obtained by the module of the analytic representation ${\cal A}[R_p(r)]$ of $R_p(r)$: $\text{envelope}[R_p(r)] = |{\cal A}[R_p(r)]|$ where ${\cal A}[R_p(r)] = R_p(r) + i {\cal H} [R_p(r)]$ and ${\cal H} [R_p(r)]$ is the Hilbert transform of $R_p(r)$. However, we decide to proceed in the following simpler manner. For each positive-energy state $p$, we determine all the local maxima $r_i$ of the absolute value of the oscillatory radial function $|R_p(r)|$ and perform a linear fit of $\ln |R_p(r_i)|$ 
\begin{equation}
\ln |R_p(r_i)| \approx \ln A_p - B_p \; r_i -C_p \ln r_i,
\label{eq:lnF}
\end{equation}
to determine the constants $\ln A_p$, $B_p$, and $C_p$.

Equations~(\ref{Gamman}) and~(\ref{eq:lnF}) define an \textit{ab initio} automatizable procedure for determining the lifetime correction to the one-electron scattering states for a given basis set. Once the values for $\gamma_p$ are determined, the complex energies $\varepsilon_p - i \gamma_p/2$ can be used in the time propagation of the Schr\"odinger equation. The presence of the finite lifetimes for the scattering states leads to a partial absorption of the wave function which simulates ionization and reduces artificial reflections of the time-dependent wave function. We stress that we do not view these lifetimes as physical lifetimes (i.e., associated with a physical resonance phenomenon), but rather as artificial lifetimes compensating the missing part of the function space due to the use of an incomplete basis set (see Ref.~\onlinecite{BerGriHie-PRA-06} for how to relate lifetimes to missing degrees of freedom). In the limit of a complete basis set, the exact continuum scattering states with $1/r$ asymptotic behavior would be obtained, and the above procedure would lead to $B_p=0$ and thus $\gamma_p=0$, as it should. On the contrary, if we use a bad basis set containing basis functions which are too much localized to represent the scattering states well, then the above procedure would lead to large values for $B_p$ and thus large values for $\gamma_p$, as we would expect.

Interestingly, for small $B_p$ (i.e., for good enough basis sets and in the lower-energy part of the continuum), we see that $\gamma_p$ is proportional to $\sqrt{2\varepsilon_p}$
\begin{eqnarray}
\gamma_p \approx 2 B_p \sqrt{2\varepsilon_p}.
\label{Gammanapprox}
\end{eqnarray}
If we set $2 B_p = 1/\tilde{d}$ for all states $p$, where $\tilde{d}$ is a single parameter to be empirically chosen, Eq.~(\ref{Gammanapprox}) reduces to the heuristic lifetime model of Klinkusch {\it et al.}~\cite{Klinkusch:2009iw}. In their reasoning, $\tilde{d}$ represents the characteristic escape length that an electron in the state $p$ with classical velocity $v_p = \sqrt{2\varepsilon_p}$ can travel during the lifetime $1/\gamma_p$. Thus, Eq.~(\ref{Gammanapprox}) can be considered as an extension of their heuristic model in which, for each scattering state and each basis set, the parameter $\tilde{d}$ is determined \textit{ab initio} by setting it to $1/(2 B_p)$, a measure of the spatial extension of the state. It seems natural indeed that the parameter $\tilde{d}$ should be different for each state. In fact, we recently proposed a slightly more flexible version of the heuristic lifetime model in which two values of $\tilde{d}$ are used for the lower-energy part and the upper-energy part of the continuum spectrum~\cite{coccia16b}. The more general formula for the inverse lifetime that we propose in Eq.~(\ref{Gamman}) corresponds to using $v_p = \sqrt{2\varepsilon_p+B_p^2}$ which indeed, as mentioned in Section~\ref{Schrodinger}, represents the velocity of the escaping electron in a complex-energy state. 

\subsection{Extension of the \textit{ab initio} lifetime correction to $N$-electron scattering states}

We discuss now the extension of our \textit{ab initio} lifetime correction from one-electron hydrogen-like systems to $N$-electron atomic and molecular systems. 

The first step of an electronic-structure calculation is usually to solve an effective one-electron mean-field Schr\"odinger equation, i.e. the Hartree-Fock (HF) or Kohn-Sham (KS) equations,
\begin{equation}
\left( -\frac{1}{2} \nabla_\b{r}^2 + v_\text{eff}(\b{r}) \right) \psi_p(\b{r}) = \varepsilon_p \psi_p(\b{r}),
\label{heffpsip}
\end{equation}
where $v_\text{eff}(\b{r})$ is an effective one-electron potential (in the case of HF, there is actually a different \textit{local} effective potential for each orbital $p$, or equivalently an unique \textit{nonlocal} effective potential, see e.g. Ref.~\onlinecite{GraKreKurGro-INC-00}). For systems with a radial effective potential $v_\text{eff}(r)$ with $r=|\b{r}|$ (i.e., atoms with spherically symmetric states), the long-range asymptotic behavior of $v_\text{eff}(r)$ as $r\to\infty$ is known: $v_\text{eff}(r) \sim -(Z-N+1)/r$ for exact KS, and $v_\text{eff}(r) \sim -(Z-N)/r$ for the virtual orbitals in HF or KS with (semi)local density-functional approximations. Therefore, the analysis of the asymptotic behavior of the radial wave function done in Section~\ref{Schrodinger} can be applied here to the radial part of each positive-energy orbital $\psi_p(\b{r})$ by just replacing $Z$ with an effective nuclear charge $Z_\text{eff} = Z-N+1$ or $Z_\text{eff}=Z-N$ (which may be zero). We can then straightforwardly apply the procedure of Section \ref{lifetimeoneelec} for each positive-energy orbital $p$, i.e perform the fit of Eq.~(\ref{eq:lnF}) and obtain the inverse lifetime $\gamma_p$ with Eq.~(\ref{Gamman}). For systems with non-spherically symmetric effective potential $v_\text{eff}(\b{r})$ (i.e., for molecules or atoms with non spherically symmetric states), for large enough $r$, any orbital $\psi_p(\b{r})$ also feels an effective potential $-Z_\text{eff}/r$ where $Z_\text{eff} = \sum_I Z_I -N +1$ or $Z_\text{eff} = \sum_I Z_I -N$ (with $Z_I$ being the charge of nucleus $I$) and $r$ can be taken as the radial coordinate around the center of mass of the system. Thus, in this case as well, we can apply the procedure of Section \ref{lifetimeoneelec} using for example the spherical average of $|\psi_p(\b{r})|^2$ around the center of mass to obtain the inverse lifetime $\gamma_p$ for each positive-energy orbital $p$.

Once the one-electron orbitals have been determined, the $N$-electron states can be determined in a second step by a many-body electronic-structure calculation, and we would like to define now lifetimes for these states. For this, we note that attributing inverse lifetimes $\gamma_p$ to the orbitals $\psi_p(\b{r})$ (without changing them), i.e. just making the replacement $\varepsilon_p \to \varepsilon_p -i\gamma_p/2$ in Eq.~(\ref{heffpsip}), formally corresponds to adding the following nonlocal one-electron complex-absorbing potential (CAP) to the Hamiltonian
\begin{equation}
v_{\text{CAP}}(\b{r},\b{r}') = -\frac{i}{2} \sum_p \gamma_p \; \psi_p^*(\b{r}') \psi_p(\b{r}),
\label{vCAP}
\end{equation}
where the sum is over the orthonormal positive-energy orbitals, or equivalently over all orbitals with the understanding that $\gamma_p = 0$ if $\varepsilon_p <0$. The CAP potential can also be conveniently expressed in second quantization
\begin{equation}
\hat{v}_{\text{CAP}} = -\frac{i}{2} \sum_p \gamma_p \; \hat{a}_p^\dag \hat{a}_p,
\label{vCAPsq}
\end{equation}
where $\hat{a}_p^\dag$ and $\hat{a}_p$ are creation and annihilation operators, respectively. We then have to include this potential in the many-body calculation. 

For example, we consider the case of the configuration interaction singles (CIS) method. In this method, the $n^\text{th}$ $N$-electron state is written as
\begin{equation}
\left| \Psi_n \right\ket= c_0 \left| \Phi_0 \right\ket + \sum_{i}^{\text{occ}}\sum_{a}^{\text{vir}} c_{i,n}^a \left| \Phi_{i}^a \right\ket,
\end{equation}
where $\left| \Phi_0 \right\ket$ is the reference HF state, $\left|\Phi_{i}^a\right\ket = \hat{a}_a^\dag \hat{a}_i \left| \Phi_0 \right\ket$ is the state obtained by the single excitation from the occupied HF orbital $i$ to the virtual HF occupied $a$, and the coefficients $c_0$ and $c_{i,n}^a$ are obtained by diagonalizing the Hamiltonian in this space. In principle, one could think of rediagonalizing the Hamiltonian including the CAP potential. A simpler approach is to just calculate the first-order correction due to the CAP potential to the energy $E_n$ of each scattering CIS state, i.e. states such that $E_n > E_0 + \text{IP}$ where $E_0$ is the ground-state energy and IP is the ionization potential. Noting that since occupied orbitals have negative energies they do not contribute in Eq.~(\ref{vCAPsq}), the action of $\hat{v}_{\text{CAP}}$ on $\left| \Phi_0 \right\ket$ is zero and $\left\bra \Phi_{i}^a \right| \hat{v}_{\text{CAP}} \left| \Phi_{j}^b \right\ket = -(i/2) \gamma_a \delta_{ab} \delta_{ij}$. We thus easily find 
\begin{equation}
\left\bra \Psi_n \right| \hat{v}_\text{CAP} \left| \Psi_n \right\ket= -\frac{i}{2} \Gamma_n \;,
\end{equation}
where $\Gamma_n$ is given by
\begin{equation}
\Gamma_n = \sum_{i}^{\text{occ}}\sum_{a}^{\text{vir}} |c_{i,n}^a|^2 \; \gamma_a,
\label{GammanCIS}
\end{equation}
with again $\gamma_a\not=0$ only if $\varepsilon_a >0$. Thus, within first order, the action of the CAP potential is to attribute inverse lifetimes $\Gamma_n$ to the scattering CIS states, i.e.
\begin{equation}
E_n \to E_n - \frac{i}{2} \Gamma_n,
\end{equation}
for $E_n > E_0 + \text{IP}$.

Equation~(\ref{GammanCIS}) exactly corresponds to the expression used in the heuristic lifetime model of Klinkusch {\it et al.}~\cite{Klinkusch:2009iw} for CIS states. We have thus provided a theoretical derivation of their expression, giving stronger support for it and allowing generalizations. For example, for configuration interactions singles doubles (CISD), it is easy to find that Eq.(\ref{GammanCIS}) now becomes
\begin{equation}
\Gamma_n = \sum_{i}^{\text{occ}}\sum_{a}^{\text{vir}} |c_{i,n}^a|^2 \; \gamma_a + \sum_{i,j}^{\text{occ}}\sum_{a,b}^{\text{vir}} |c_{ij,n}^{ab}|^2 \; (\gamma_a + \gamma_b),
\label{GammanCISD}
\end{equation}
and so on. Alternatively, one could use directly the CAP potential of Eq.~(\ref{vCAP}) or (\ref{vCAPsq}) in time-dependent methods such as TDDFT or TDHF.

\section{Computational details}
\label{details}

We test our \textit{ab initio} lifetime correction on the H and He atoms. We start with standard Gaussian-type correlation-consistent polarized valence-triple-zeta Dunning basis sets~\cite{Dun-JCP-89}, $n$-fold augmented with diffuse basis functions to describe Rydberg states~\cite{lupp+13jcp}, denominated by $n$-aug-cc-pVTZ. For the atoms considered, these basis sets contain s, p, and d basis functions. For each angular momentum, we then add $m$ Gaussian-type functions adjusted to represent low-lying continuum states, as proposed by Kaufmann {\it et al.}~\cite{kauf+89physb} and used in Refs.~\onlinecite{coccia16a,coccia16b}. The resulting basis sets are referred to as $n$-aug-cc-pVTZ+$m$K where K stands for ``Kaufmann''. Specifically, we consider $n=6$ or $n=8$ and $m=3$ or $m=8$ for the H atom, and $n=6$ and $m=7$ for the He atom. Note that $m=8$ and $m=7$ for H and He, respectively, are the largest numbers of Kaufmann functions that we have been able to use before running into linear-dependency problems. We have recently extensively studied the convergence of the HHG spectrum of the H atom with such basis sets and found that the 6-aug-cc-pVTZ+8K basis set with a two-parameter heuristic lifetime model already gives a HHG spectrum in good agreement with the reference grid-based one (for laser intensities up to $10^{14}$ W/cm$^2$)~\cite{coccia16b}.

Using a development version of the \textsc{Molpro} software package~\cite{MOLPRO_brief}, we perform a Hartree-Fock calculation to obtain the orbitals with these different basis sets. To obtain the inverse lifetime $\gamma_p$ for each positive-energy orbital $p$, we numerically determine the local maxima of the absolute value of the radial part of the orbital using a spatial grid with step 0.05 bohr extending from $r_\text{min} = 0.05$ bohr to $r_\text{max} = 2/\sqrt{\alpha_{\text{min,s}}}$ where $\alpha_{\text{min,s}}$ is the exponent of the most diffuse s-function in the basis set~\cite{white15}. With the largest basis sets used, we have $r_\text{max} = 419$ bohr with the 6-aug-cc-pVTZ+8K basis set for the H atom, and $r_\text{max} = 295$ bohr with the 6-aug-cc-pVTZ+7K basis set for the He atom. We then perform the fit in Eq.~(\ref{eq:lnF}). 

We perform a CIS calculation (for the H atom, this is of course identical to HF) to obtain CIS total energies $E_n$ and coefficients $c_{i,n}^a$, as well as transition moments, and calculate the CIS inverse lifetimes $\Gamma_n$ according to Eq.~(\ref{GammanCIS}) for the states $n$ such that $E_n > E_0 + \text{IP}$ where $E_0$ is the HF ground-state energy. For the ionization potential, we take $\text{IP} = - \varepsilon_\text{HOMO}$ calculated with the considered basis set, giving $\text{IP}=0.5$ Ha for H and $\text{IP}=0.918$ Ha for He. 

To test the obtained lifetimes, we calculate HHG spectra (in the dipole form) induced by a $\cos^2$-shape laser electric field by performing TDCIS calculations with the real-time propagation code \textsc{Light}~\cite{lupp+13jcp} using a time step $\Delta t$ = 2.42 as (0.1 a.u.) and the same set up as in Ref.~\onlinecite{coccia16b}. Specifically, for H we use a laser intensity of $I=10^{14}$ W/cm$^{2}$ with a wavelength of $\lambda_0= 800$ nm, and for He we use a laser intensity of $I=5 \times 10^{14}$ W/cm$^{2}$ with a wavelength of $\lambda_0= 456$ nm \cite{ding+11jcp}. In both cases, the time propagation has been carried out for 20 optical cycles.

\section{Results}
\label{results}

\subsection{Hydrogen atom}

\begin{figure}
\begin{center}
\includegraphics[scale=0.35,angle=-90]{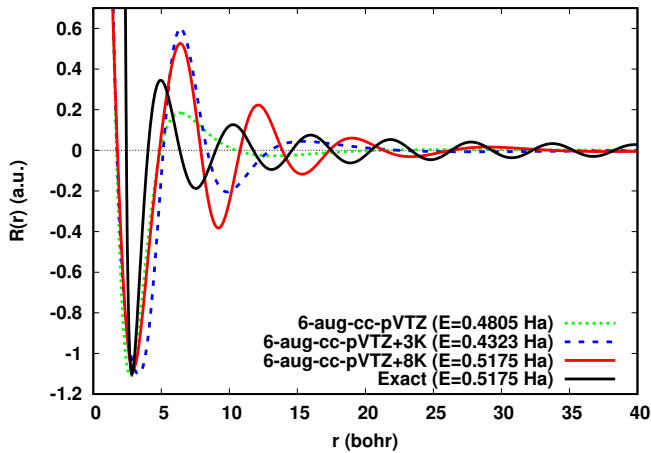}
\caption{Comparison of the radial wave function $R(r)$ of the exact s-symmetry $(\ell=0)$ scattering state of energy $0.5175$ Ha of the H atom with the radial wave function of the approximate scattering state of closest energy obtained with the 6-aug-cc-pVTZ, 6-aug-cc-pVTZ+3K, or 6-aug-cc-pVTZ+8K basis set (energy of 0.4805 Ha, 0.4323 Ha, and 0.5175 Ha, respectively). The curves are normalized such that they (approximately) have the same amplitude at the first minimum.
\label{fig:wf}} 
\end{center}
\end{figure} 

We start by showing the typical radial wave functions of scattering states of the H atom that we obtain with Gaussian-type basis sets. In Fig.~\ref{fig:wf}, we compare the radial wave function $R(r)$ of the exact s-symmetry scattering state of energy $0.5175$ Ha with the radial wave function of the approximate scattering state of closest energy obtained with the 6-aug-cc-pVTZ, 6-aug-cc-pVTZ+3K, or 6-aug-cc-pVTZ+8K basis set (0.4805 Ha, 0.4323 Ha, and 0.5175 Ha, respectively). As observed in Ref.~\onlinecite{coccia16b}, the 6-aug-cc-pVTZ basis set does not reproduce the long-range oscillatory behavior of the exact scattering wave function. The situation is improved when adding Kaufmann functions, i.e. with the 6-aug-cc-pVTZ+3K and 6-aug-cc-pVTZ+8K basis sets. The more Kaufmann functions are added, the more long-range oscillations are obtained in the radial wave functions. However, even with the 6-aug-cc-pVTZ+8K basis set, the amplitude of these oscillations decay much too fast at long distance in comparison with the exact $1/r$ behavior. It is not easy to continue to improve the 6-aug-cc-pVTZ+8K basis set by adding more and more Kaufmann functions because of linear dependencies. Instead, we will compensate for this wrong asymptotic behavior using our \textit{ab initio} lifetime correction.

%

\begin{table}
  \caption{\label{tab:1} Evolution of the values of the parameters $B_{p}$ and $C_{p}$, and of the coefficients of determination $R^2$ for the fit in Eq.~(\ref{eq:lnF}) when removing a number of maxima $r_i$ that are the largest distances (indicated with the number of total maxima $N_{\text{max}}$ and the position of the last maximum $r_\text{lastmax}$ used for the fit) for the s, p, and, d states of the H atom considered in Fig.~\ref{fit}. The 6-aug-cc-pVTZ+8K basis set has been used. There are no maxima beyond $r = 228.7$ bohr.}
   \begin{tabular}{c  c  c  c  c  }
   \hline\hline
   $N_{\text{max}}$ & \phantom{x} $r_\text{lastmax}$ (bohr) \phantom{x}   & $B_{p}$ (bohr$^{-1}$)& $C_{p}$ & $R^{2}$ \\
   \hline\\[-0.2cm]
    \multicolumn{5}{c}{s state at $\varepsilon = 0.343$ Ha}\\
     11  & 228.7  &   0.013   & 2.022  &  0.97   \\
     10  & 111.6  &  0.037    & 1.436  &  0.97  \\
     9   & 59.5   &   0.087   & 0.633  &  0.99  \\
     8   & 37.9   &    0.130  & 0.106  &  0.99 \\[0.1cm]
    \multicolumn{5}{c}{p state at $\varepsilon = 0.306$ Ha}\\
     9   & 112.1 &   0.043    & 1.220  & 0.96 \\
     8   & 57.9  &   0.096    & 0.518  &  0.98 \\
     7   & 34.0  &   0.150    &  -5$\times10^{-4}$ & 0.99 \\[0.1cm]
    \multicolumn{5}{c}{d state at $\varepsilon = 0.371$ Ha}\\
     8   & 102.4 &    0.045   &  1.495 &  0.97  \\
     7   & 50.0  &    0.120   &  0.406  &   0.98 \\
   \hline\hline
  \end{tabular}
\end{table}

\begin{figure}
\begin{center}
\includegraphics[scale=0.3,angle=-90]{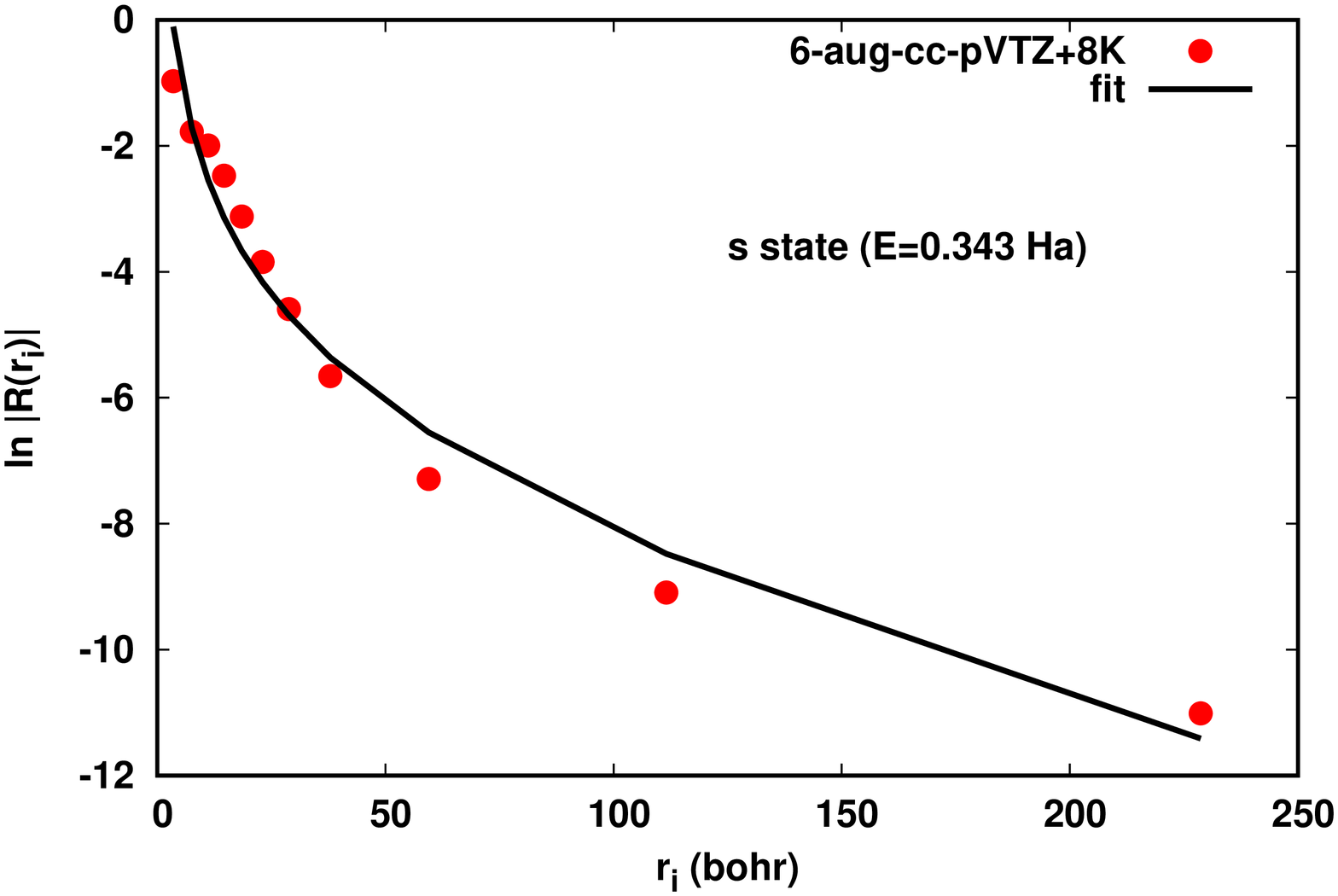}
\includegraphics[scale=0.3,angle=-90]{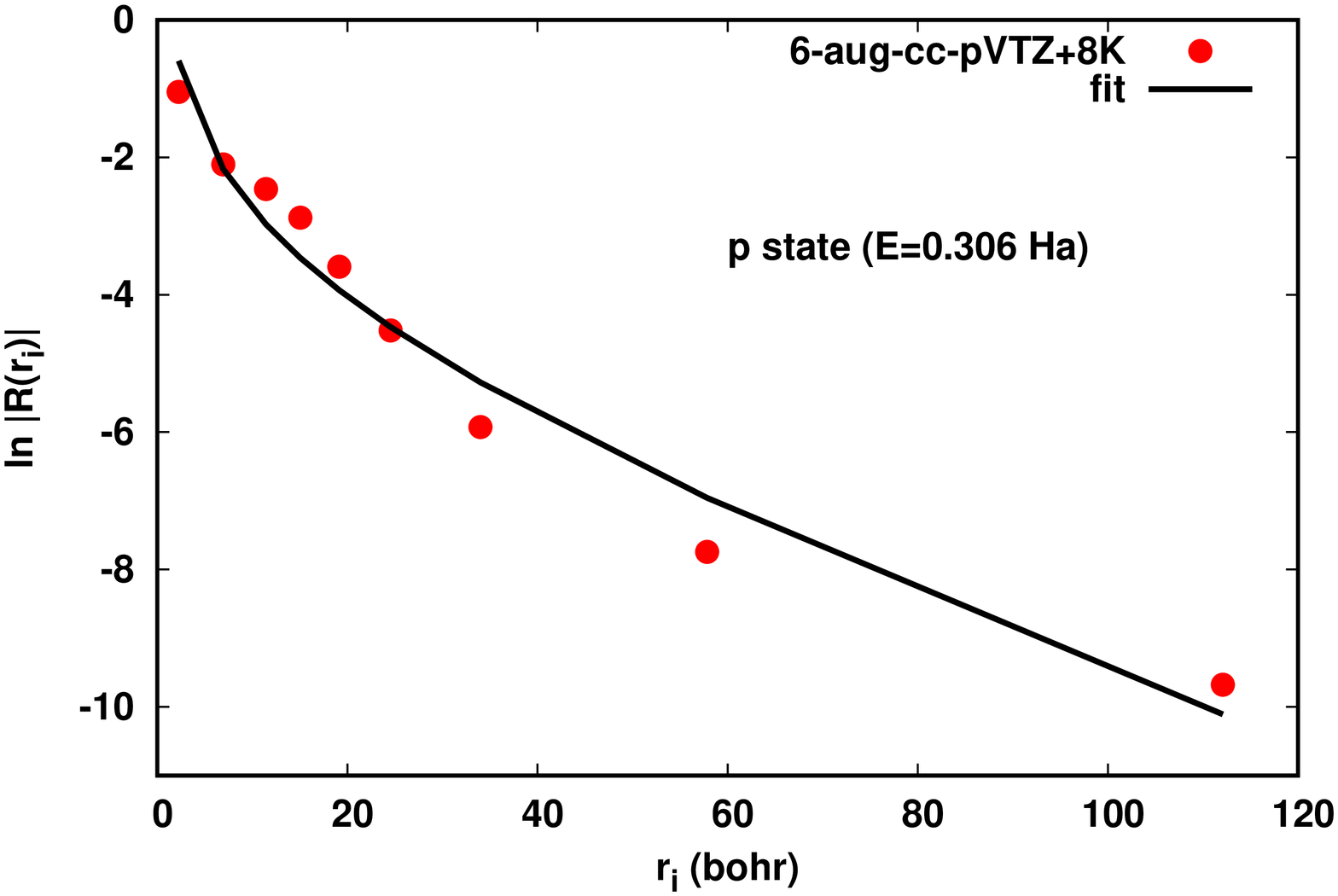}
\includegraphics[scale=0.3,angle=-90]{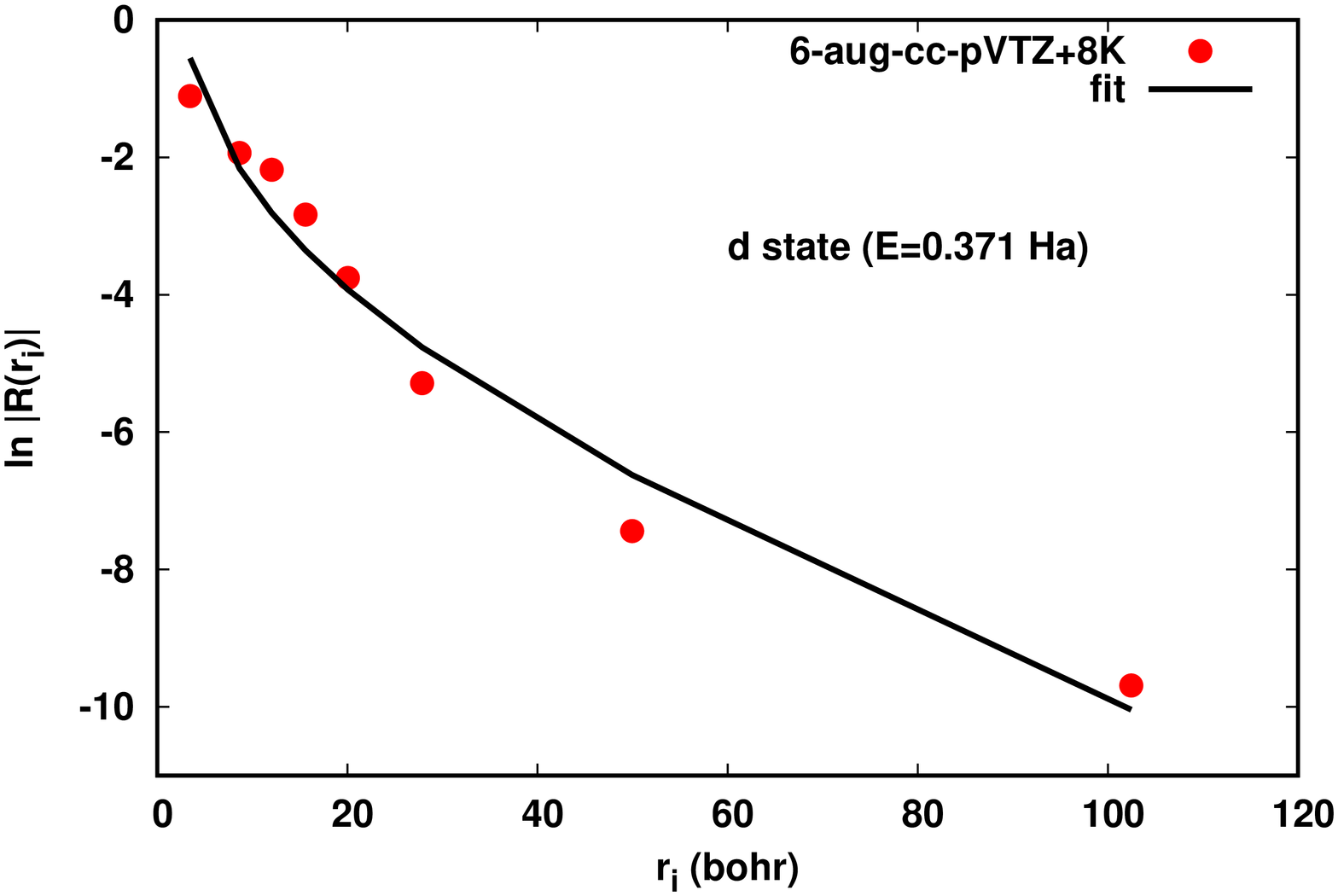}
\caption{Fit to the logarithmic formula of Eq.~(\ref{eq:lnF}) of the envelope of the s, p, and d radial wave functions $R(r)$ of the H atom with positive energies of 0.343 Ha, 0.306 Ha, and 0.371 Ha, respectively, calculated with the 6-aug-cc-pVTZ+8K basis set. The points defining the envelope are defined as the maxima $r_i$ of $|R(r)|$. The coefficients of determination $R^2$ of the fit are 0.97, 0.96 and 0.96, respectively.
\label{fit}} 
\end{center}
\end{figure}

We consider now the fit of the envelope of the radial wave functions with the logarithmic expression of Eq.~(\ref{eq:lnF}). In Fig.~\ref{fit}, we compare the values $\ln|R_p(r_i)|$ where $r_i$ are the maxima of $|R_p(r)|$ with the fitted curve $\ln A_p - B_p \; r -C_p \ln r$ for s, p, and d scattering wave functions of similar energies (0.343 Ha, 0.306 Ha, and 0.371 Ha, respectively~\cite{lifetimes-note3}) calculated with the 6-aug-cc-pVTZ+8K basis set. As mentioned in the Computational details, the fit was performed within the radial window $[r_\text{min} = 0.05; r_\text{max} = 419]$ bohr in which there are 11, 9, and 8 maxima of $|R_p(r)|$ for the s, p, and d states considered, respectively. In Table~\ref{tab:1}, we also report the values of the parameters $B_p$ and $C_p$, and the coefficients of determination $R^{2}$ of the fits, obtained with different numbers of maxima included, corresponding to using smaller values of $r_\text{max}$. The quality of the fit is satisfactory with $R^2 \geq 0.96$ in all cases, but the value of $B_p$ appears to be quite sensitive to the number of maxima included and increases significantly when reducing $r_\text{max}$. The value $r_\text{max} = 2/\sqrt{\alpha_{\text{min,s}}} = 419$ bohr chosen in this work thus gives the smallest value of $B_p$ and consequently the smallest value of the inverse lifetime $\gamma_p$. This is in a sense a ``safe'' choice since it minimizes the lifetime correction.


\begin{figure}
\begin{center}
\includegraphics[scale=0.3,angle=-90]{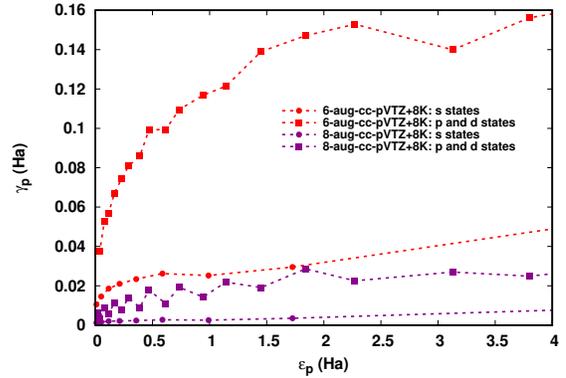}
\caption{Inverse lifetimes $\gamma_p$ obtained with the fit from Eq.~(\ref{Gamman}) as a function of the orbital energies $\varepsilon_p$ for the H atom with the 6-aug-cc-pVTZ+8K basis and 8-aug-cc-pVTZ+8K basis sets for s scattering states and p and d scattering states.
\label{fig:comp}} 
\end{center}
\end{figure} 

Let us discuss now the inverse lifetimes $\gamma_p$ for each scattering state obtained with the fit from Eq.~(\ref{Gamman}). In Fig.~\ref{fig:comp} we show $\gamma_p$ obtained with 6-aug-cc-pVTZ+8K and 8-aug-cc-pVTZ+8K basis sets as a function of the orbital energies $\varepsilon_p$. Consider first the 6-aug-cc-pVTZ+8K basis set. The inverse lifetimes for the s scattering states and the ones for the p and d scattering states both roughly follow a $\sqrt{\varepsilon_p}$ trend. What is striking is that the inverse lifetimes of the p and d scattering states are much larger than the inverse lifetimes of the s scattering states. Since the inverse lifetimes should be zero in the limit of a complete basis set, this must mean that the 6-aug-cc-pVTZ+8K basis set is much worse for the p and d scattering states in comparison to the s scattering states. Consider now the 8-aug-cc-pVTZ+8K basis set. As expected, with this improved basis set containing more diffuse functions, we obtain much smaller inverse lifetimes for all scattering states. However, the inverse lifetimes for p and d scattering states with this basis set are still larger than the ones for s scattering states. We thus conclude that both basis sets are unbalanced in the description the s scattering states and the p and d scattering states. It is a nice feature of our \textit{ab initio} lifetime correction that it clearly reveals the imbalance of the basis set for scattering states of different angular momenta.

\begin{figure}
\begin{center}
\includegraphics[scale=0.3,angle=-90]{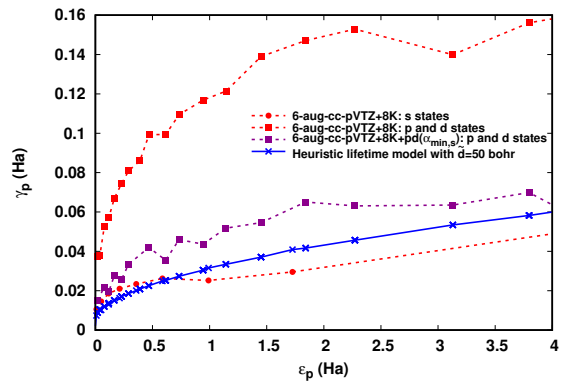}
\caption{Inverse lifetimes $\gamma_p$ obtained with the fit from Eq.~(\ref{Gamman}) for the H atom with the 6-aug-cc-pVTZ+8K and 6-aug-cc-pVTZ+8K+pd($\alpha_{\min,\text{s}}$) basis sets for s scattering states and p and d scattering states. The 6-aug-cc-pVTZ+8K+pd($\alpha_{\min,\text{s}}$) basis set is obtained from the 6-aug-cc-pVTZ+8K basis set by adding p and d basis functions with the $\alpha_{\min,\text{s}}$ exponent. The inverse lifetimes obtained from the heuristic lifetime model~\cite{Klinkusch:2009iw} with the 6-aug-cc-pVTZ+8K+pd($\alpha_{\min,\text{s}}$) basis set and $\tilde{d}=50$ bohr are also shown.
\label{fig:basis}} 
\end{center}
\end{figure} 

This better description of the s scattering states than the p and d scattering states may be explained by the fact that the most diffuse basis functions of the $n$-aug-cc-pVTZ+8K basis sets are of s symmetry. To confirm this hypothesis, we have constructed a new basis set starting from the 6-aug-cc-pVTZ+8K basis set and adding p and d basis functions with the smallest exponent of the s basis functions in this basis, which is $\alpha_{\min,\text{s}} = 2.28\times10^{-5}$ bohr$^{-2}$. In Fig.~\ref{fig:basis}, it is seen that the resulting basis set, denoted by 6-aug-cc-pVTZ+8K+pd$(\alpha_{\min,\text{s}})$, gives of course the same inverse lifetimes for the s scattering states, but much smaller inverse lifetimes for the p and d scattering states which are now comparable to the inverse lifetimes of the s scattering states. The 6-aug-cc-pVTZ+8K+pd$(\alpha_{\min,\text{s}})$ basis set is thus a more balanced basis set. In Fig.~\ref{fig:basis}, we also show the inverse lifetimes obtained from the heuristic lifetime model of Klinkusch {\it et al.}~\cite{Klinkusch:2009iw}, i.e. $\gamma_p = \sqrt{2\varepsilon_p}/\tilde{d}$, with the 6-aug-cc-pVTZ+8K+pd$(\alpha_{\min,\text{s}})$ basis set and the value of $\tilde{d}=50$ bohr. This value of $\tilde{d}$ was empirically found in a previous work~\cite{coccia16b} to give a good HHG spectrum of the H atom with the aug-cc-pVTZ+8K basis set, in good agreement with the reference HHG spectrum obtained from grid calculations. Clearly, for the s scattering states, the inverse lifetimes obtained from the heuristic lifetime model with this value of $\tilde{d}$ are quite similar to the inverse lifetimes determined \textit{ab initio} in the present work. For the p and d scattering states, the heuristic lifetime model gives inverse lifetimes that are a bit smaller than the \textit{ab initio} inverse lifetimes obtained with the 6-aug-cc-pVTZ+8K+pd$(\alpha_{\min,\text{s}})$ basis set. Therefore, we can consider that our \textit{ab initio} lifetime correction provides a first-principle justification for the value of $\tilde{d}$ empirically chosen in Ref.~\onlinecite{coccia16b}.

\begin{figure}
\begin{center}
\includegraphics[scale=0.35,angle=-90]{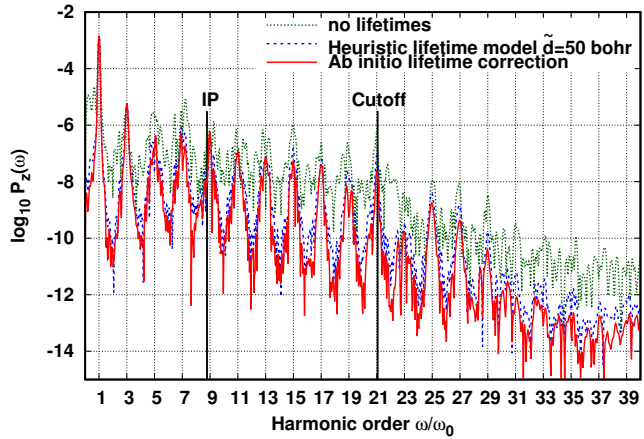}
\caption{HHG spectra of the H atom with laser intensity $I = 10^{14}$ W/cm$^2$ calculated with the 6-aug-cc-pVTZ+8K+pd($\alpha_{\min,\text{s}}$) basis set using either no lifetimes, the lifetimes from the heuristic lifetime model~\cite{Klinkusch:2009iw} using $\tilde{d}=50$ bohr or the lifetimes from our {\it ab initio} lifetime correction. The vertical lines correspond to the value of the IP and the three-step model energy cutoff~\cite{cork93prl,lewe+pra94}.
\label{fig:hhg_h}} 
\end{center}
\end{figure} 
  
Finally, we test our \textit{ab initio} lifetime correction for calculating the HHG spectrum of the H atom with a laser intensity of $I = 10^{14}$ W/cm$^{2}$ using the 6-aug-cc-pVTZ+8K+pd($\alpha_{\min,\text{s}}$) basis set. We show the obtained spectrum in Fig.~\ref{fig:hhg_h} and compare it to the HHG spectra calculated using either no lifetimes or lifetimes from the heuristic lifetime model with $\tilde{d}=50$ bohr. All the spectra present roughly the expected aspect of an atomic HHG spectrum: a first intense peak at $\omega/\omega_0=1$, followed by a plateau of peaks at the odd harmonic orders until a cutoff value beyond which the intensity of the peaks rapidly decreases. The spectrum obtained with no lifetimes is however very noisy, the signal not decreasing very much between the peaks. Introducing the lifetimes results in much clearer spectra with lower background and sharper peaks. The spectrum obtained with the lifetimes from the \textit{ab initio} procedure and the one obtained the lifetimes from the heuristic model are very similar to each other, with the \textit{ab initio} lifetimes giving a slightly lower background (less than one unit on the logarithmic scale). This test thus confirms the usefulness of introducing lifetimes, and confirm that the heuristic lifetime model can be replaced by our \textit{ab initio} lifetime correction.

\subsection{Helium atom}

\begin{figure*}
\begin{center}
\includegraphics[scale=0.35,angle=-90]{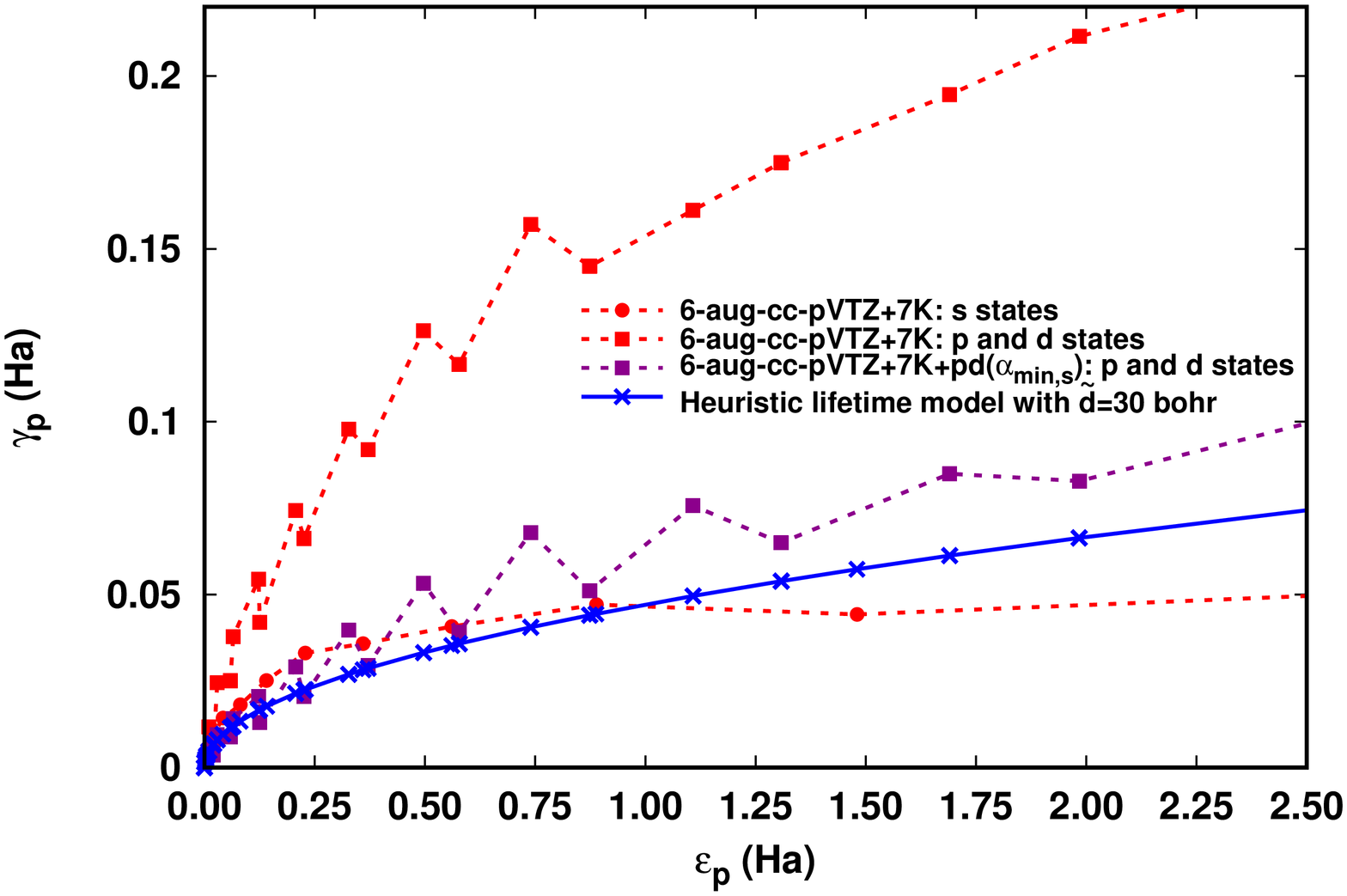}
\includegraphics[scale=0.35,angle=-90]{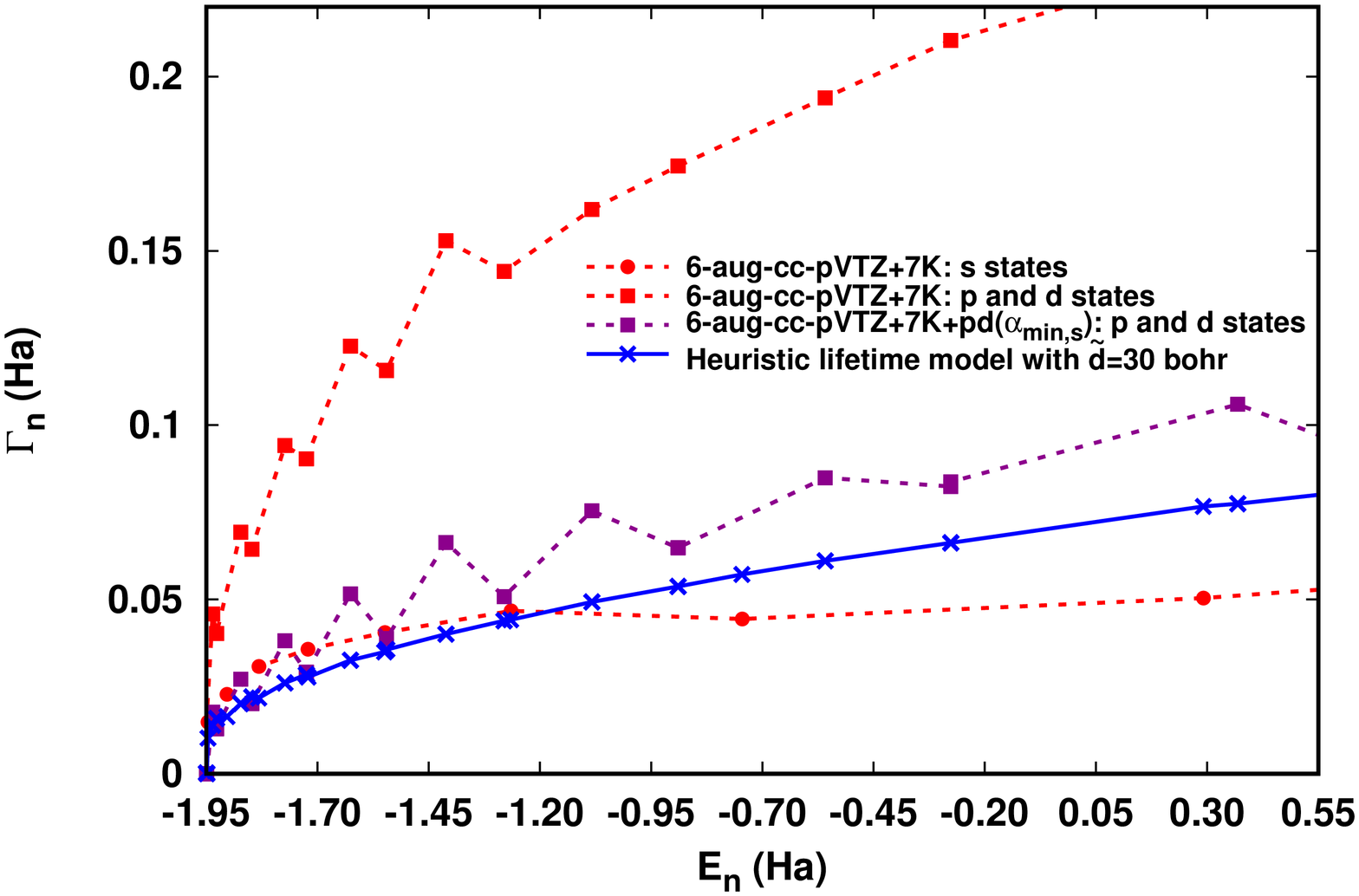}
\caption{
One-electron inverse lifetimes $\gamma_p$ as a function of the orbital energies $\varepsilon_p$ (left) and two-electron CIS inverse lifetimes $\Gamma_n$ as a function of the CIS total energies $E_n$ (right) for the He atom with the 6-aug-cc-pVTZ+7K and 6-aug-cc-pVTZ+7K+pd($\alpha_{\min,\text{s}}$) basis sets for s scattering states and p and d scattering states. The inverse lifetimes obtained from the heuristic lifetime model~\cite{Klinkusch:2009iw} with the 6-aug-cc-pVTZ+7K+pd($\alpha_{\min,\text{s}}$) basis set and $\tilde{d}=30$ bohr are also shown.
\label{fig:bas_he}} 
\end{center}
\end{figure*} 

We apply now our \textit{ab initio} lifetime correction to the He atom, as a first test on a system with more than one electron. We consider first the one-electron inverse lifetimes $\gamma_p$ as a function of the orbital energies $\varepsilon_p$ which are reported in the left panel of Fig.~\ref{fig:bas_he} using the 6-aug-cc-pVTZ+7K and 6-aug-cc-pVTZ+7K+pd($\alpha_{\min,\text{s}}$) basis sets. Similarly as for the H atom, the basis 6-aug-cc-pVTZ+7K+pd($\alpha_{\min,\text{s}}$) is constructed from the basis 6-aug-cc-pVTZ+7K by adding p and d basis functions with the smallest s-basis function exponent. With the 6-aug-cc-pVTZ+7K basis set, the inverse lifetimes of the s scattering states are much smaller than the ones of the p and d scattering states. As for the H atom, the use of the 6-aug-cc-pVTZ+7K+pd($\alpha_{\min,\text{s}}$) basis set gives a more balanced description of all the scattering states. The obtained inverse lifetimes follow a similar trend as the one observed for the H atom with a similar basis set, but tend to be a bit larger. As a consequence, if we want to roughly reproduce these \textit{ab initio} lifetimes with the heuristic lifetime model, we need to choose a smaller value of the parameter: $\tilde{d}=30$ bohr. We consider now the corresponding two-electron CIS inverse lifetimes $\Gamma_n$ [Eq.~(\ref{GammanCIS})] as a function of the CIS total energies $E_n$, reported in the right panel of Fig.~\ref{fig:bas_he}. The CIS inverse lifetimes are overall quite similar to the one-electron inverse lifetimes. The most important difference is that, just above the continuum threshold ($\varepsilon_p \gtrsim 0$ or $E_n \gtrsim E_0+\text{IP}$), the density of two-electron CIS states is higher than the density of one-electron HF states, and the CIS inverse lifetimes are significantly larger than the one-electron inverse lifetimes.

\begin{figure}
\begin{center}
\includegraphics[scale=0.35,angle=-90]{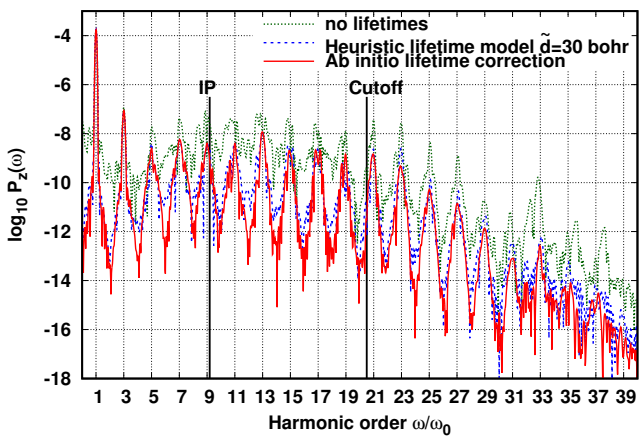}
\caption{HHG spectra of the He atom with laser intensity $I = 5\times 10^{14}$ W/cm$^2$ calculated with the 6-aug-cc-pVTZ+7K+pd($\alpha_{\min,\text{s}}$) basis set using either no lifetimes, the lifetimes from the heuristic lifetime model~\cite{Klinkusch:2009iw} using $\tilde{d}=30$ bohr or the lifetimes from our {\it ab initio} lifetime correction. The vertical lines correspond to the value of the IP and the three-step model energy cutoff~\cite{cork93prl,lewe+pra94}.
\label{fig:hhg_he}} 
\end{center}
\end{figure}

Finally, we test our \textit{ab initio} lifetime correction for calculating the HHG spectrum of the He atom with a laser intensity of $I = 5 \times 10^{14}$ W/cm$^{2}$ using the 6-aug-cc-pVTZ+7K+pd($\alpha_{\min,\text{s}}$) basis set. Since the study of the effect of electronic correlation on the HHG spectrum~\cite{shi13} is beyond the scope of this work, we still use TDCIS even though the He atom has two electrons, i.e. we neglect double excitations. We show the obtained spectrum in Fig.~\ref{fig:hhg_he} and compare it to the HHG spectra calculated using either no lifetimes or lifetimes from the heuristic lifetime model with $\tilde{d}=30$ bohr. As for the H atom, the spectrum obtained with no lifetimes is very noisy, whereas the spectra obtained with the lifetimes are much clearer. Using the \textit{ab initio} lifetimes gives a slightly lower background than using the lifetimes from the heuristic lifetime model. This test thus confirms the applicability of our \textit{ab initio} lifetime correction to two-electron systems.

\section{Conclusions}
\label{conclusion}

We have developed a method for obtaining effective lifetimes of scattering electronic states for avoiding the artificially confinement of the wave function due to the use of incomplete basis sets in time-dependent electronic-structure calculations. In this method, using a fitting procedure, the lifetimes are systematically extracted from the spatial asymptotic decay of the approximate scattering wave functions obtained with a given basis set. The main qualities of this method are that (1) it is based on a rigorous theoretical analysis, (2) it does not involve any empirical parameters, (3) it is adapted to each particular basis set used. Interestingly, the method can be considered as an \textit{ab initio} version of the heuristic lifetime model of Klinkusch {\it et al.}~\cite{Klinkusch:2009iw}.

As first tests of our method, we have considered the H and He atoms using Gaussian-type basis sets. We have shown that reasonable lifetimes adapted to the basis set are obtained. In particular, the inverse lifetimes correctly decrease when the size of the basis set is increased. Moreover, the obtained lifetimes revealed an unbalanced description of the scattering states of different angular momentum with the standard basis sets used, which we exploited to construct more balanced basis sets. Therefore, the method is useful to diagnose the quality of a basis set for describing scattering states. Finally, the obtained lifetimes have been shown to lead to much clearer HHG spectra (i.e., with a lower background and better resolved peaks) in time-dependent calculations.

Future work includes testing the method on larger systems including molecules, calculating other properties than HHG spectra, and possibly using different types of basis sets. We believe that our approach could help adapting quantum-chemistry methods for the study of electron dynamics induced by high-intensity laser in atoms and molecules.

\vspace{1cm}
\section*{Acknowledgements}

This work was supported by the LabEx MiChem part of French state funds managed by the ANR within the Investissements d'Avenir programme under reference ANR-11-IDEX-0004-02. EC thanks L. Guidoni for support.


\begin{thebibliography}{10}
\bibitem{lifetimes-note2}
The general solution of the Schr\"odinger equation given in
  Eqs.~(\ref{solR})-(\ref{solR2}) is valid for $Z\not=0$. For $Z=0$, $R_1(r)$
  in Eq.~(\ref{solR1}) becomes identically zero, and should be replaced by a
  function which has the same expression as $R_2(r)$ with the substitution
  $\sqrt{-2E} \to -\sqrt{-2E}$. The continuum scattering states are then
  obtained as a linear combination of these two functions. However, our
  analysis of complex-energy states remains unchanged. Thus, our procedure for
  determining the lifetimes is still valid for $Z=0$.
\bibitem{lifetimes-note1}
For $\ell=0$, $R_2(r)$ only diverges as $1/r$ at $r=0$ which is normalizable,
  i.e. $\int_0^\infty \text{d}r r^2 |R_2(r)|^2 < \infty$. However, even in this
  case, the expectation value of the Hamiltonian operator is infinite. For
  $\ell \geq 1$, $R_2(r)$ is not even normalizable.
\bibitem{lifetimes-note3}
Because we do not impose spherical symmetry in our calculations, we do not have
  exactly degenerate s, p, and d states.
\end{thebibliography}
\end{document}